\documentclass[conference]{IEEEtran}

\usepackage{graphicx}
\usepackage{amsmath}
\usepackage{cite}
\usepackage{multirow}
\usepackage{subfigure}
\usepackage{stfloats} %put equation on top
\usepackage{amssymb}

\usepackage{amsthm}
\theoremstyle{definition} \newtheorem{lemma}{Lemma}

\begin{document}

\title{Angle-Domain Doppler Pre-Compensation for High-Mobility OFDM Uplink with a Massive ULA}

\author{\IEEEauthorblockN{Wei Guo\IEEEauthorrefmark{1}, Weile Zhang\IEEEauthorrefmark{1}, Pengcheng Mu\IEEEauthorrefmark{1}, Feifei Gao\IEEEauthorrefmark{2} and Bobin Yao\IEEEauthorrefmark{3}}
\IEEEauthorblockA{\IEEEauthorrefmark{1}MOE Key Lab for Intelligent Networks and Network Security, Xi'an Jiaotong University, Xi'an, China}
\IEEEauthorblockA{\IEEEauthorrefmark{2}Tsinghua National Laboratory for Information Science and Technology, Beijing, China}
\IEEEauthorblockA{\IEEEauthorrefmark{3}School of Electronic and Control Engineering, Chang'an University, Xi'an, China}
%\IEEEauthorblockA{guowei@stu.xjtu.edu.cn, wlzhang@mail.xjtu.edu.cn, pcmu@mail.xjtu.edu.cn, feifeigao@ieee.org}
}
%\thanks{This work is partially supported by the NSFC under grant No. 61172093, 61302066, \& 61302069, the Research Foundation for Doctoral Program of Higher Education of China under Grant No. 20130201120021, the Natural Science Basic Research Plan in Shaanxi Province of China under Grants No. 2014JQ8303 \& No. 2014JM2-6124.}
\maketitle

\begin{abstract}
In this paper, we propose a Doppler pre-compensation scheme for high-mobility orthogonal frequency division multiplexing (OFDM) uplink, where a high-speed terminal transmits signals to the base station (BS). Considering that the time-varying multipath channel consists of multiple Doppler frequency offsets (DFOs) with different angle of departures (AoDs), we propose to perform DFO pre-compensation at the transmitter with a large-scale uniform linear array (ULA). The transmitted signal passes through a beamforming network with high-spatial resolution to produce multiple parallel branches. Each branch transmits signal towards one direction thus it is affected by one dominant DFO when passing over the time-varying channel. Therefore, we can compensate the DFO for each branch at the transmitter previously. Theoretical analysis for the Doppler spread of the equivalent uplink channel is also conducted. It is found that when the number of transmit antennas is sufficiently large, the time-variation of channel can be efficiently suppressed. Therefore, the performance will not degrade significantly if applying the conventional time-invariant channel estimation and equalization methods at the receiver. Simulation results are provided to verify the proposed scheme.

\end{abstract}

%\begin{IEEEkeywords}
%OFDM, time-varying channels, high-mobility, carrier frequency offset (CFO), Doppler shifts, oscillator frequency offset (OFO), large-scale antenna array.
%\end{IEEEkeywords}

\section{Introduction}
% As one of the key scenarios in fifth generation (5G) communications, high-mobility communications have attracted considerable attentions. High-mobility communications have attracted considerable attentions during the past few years. To meet the requirements of high data rates,,Zhu13TDD
Orthogonal frequency division multiplexing (OFDM) has become a dominant technique for high-mobility wireless communications~\cite{Wu16A}. It can provide high spectral efficiency and is robust to the frequency-selective channels~\cite{Hwang09OFDM}. However, OFDM is sensitive to the carrier frequency offsets (CFOs), such as oscillator frequency offset (OFO)~\cite{Hwang09OFDM} and Doppler frequency offset (DFO)~\cite{Wu16A}. Especially, the Doppler spread caused by the high-speed of the moving terminal makes the channel vary with time rapidly and thus introduces inter-carrier interference (ICI) that could significantly deteriorates the link performance. The Doppler compensation is challenging since there are multiple DFOs associated with various multipaths. %are mixed at the receiver.

During the past few years, how to combat the Doppler shifts has been widely studied in the literature. Since multiple DFOs are mixed together, it is hard to estimate the DFOs directly and thus, the composite time-varying channel estimation has been intensively investigated. For example, the basic expansion model (BEM) has been used to characterize the time-variations of channel with reduced number of parameters~\cite{Hijazi09Polynomial,Tang07Pilot}. In~\cite{Hijazi09Polynomial}, the time-varying channel is characterized by a polynomial model, while in~\cite{Tang07Pilot}, the complex exponential BEM (CE-BEM) is exploited with the accurate knowledge of the maximum Doppler shift. For sparse channel with limited multipaths between the base station (BS) and the moving terminal, the compressed sensing (CS) based channel estimation methods are beneficial~\cite{Berger10Application}. %,Bajwa10Compressed
%To avoid the CFOs estimation in high-speed communications, some approaches directly estimate the channel to mitigate the ICI caused by the CFOs. However, the time-varying channel estimation is a challenge due to the vast number of parameters to be estimated.
%In the BEM based schemes, the estimation accuracy depends on the order and the basis of the expansion model. In high Doppler spread scenarios, a large number of basis are needed which will also increase the computation complexity.
%However, this scenario can not reflect the variety of channel. In this paper, we will consider a more general time-varying multipath channel with rich reflectors around the moving terminals.
These methods estimate the time-varying channel at either time or frequency domains. However, they cannot estimate and compensate multiple DFOs directly since the DFOs are mixed together and single antenna techniques are employed.

Considering that the multiple DFOs are associated with the angle of arrivals (AoAs) or angle of departures (AoDs) for multipaths, some pioneer works have been done to mitigate the effect of the DFOs from angle domain via small-scale antenna arrays~\cite{Zhang11Multiple,Yang13Beamforming,Guo13Multiple}.
%In~\cite{Klenner07Doppler}, the signal is received by sectorial antennas to reduce the range of Doppler spread.
%If the signal is received by sectorial antennas~\cite{Klenner07Doppler}, the range of Doppler spread reduces, and the ICI can be partially compensated by regarding the Doppler spread as an equivalent CFO.
In~\cite{Zhang11Multiple}, the DFOs associated with the dominant multipaths in a sparse channel are compensated separately with the knowledge of the maximum Doppler shift and the AoAs. Similarly, the DFOs for the line-of-sight (LoS) from different BSs are compensated via AoAs estimation in~\cite{Yang13Beamforming}. In~\cite{Guo13Multiple}, a small-scale antenna array is also applied to mitigate the DFOs previously with the knowledge of the AoDs in the uplink transmission. %Note that these works usually require accurate angle information for the multipaths.

%However, when the receiver or the transmitter moves at a high speed, the effect of the various Doppler shifts associated with various multipaths becomes very serious~\cite{Zhu13TDD}. In this case, the existence of multiple DFOs accompanying single OFO greatly increases the complexity of CFOs estimation and compensation.
%To avoid the CFOs estimation in high-speed communications, some approaches directly estimate the channel to mitigate the ICI caused by the CFOs. However, the time-varying channel estimation is a challenge due to the vast number of parameters to be estimated.

%However, when the receiver or the transmitter moves at a high speed, the effect of the various Doppler shifts associated with various multipaths becomes very serious~\cite{Ahmed05Parameter,Yu07Iterative}.
%However, the scenarios of multiple DFOs associated with various multipaths are less investigated. These scenarios could arise when either the receiver or the transmitter moves at a very high speed resulting in various Doppler shifts for paths with different angle of arrivals (AoAs)~\cite{Ahmed05Parameter,Yu07Iterative}.
%In the case of single CFO, such as the OFO, the estimated CFO can be directly compensated by the phase-locked loop (PLL). However, the existence of multiple DFOs accompanying single OFO greatly increases the complexity of CFOs estimation and compensation.

Most of these works based on small-scale antenna array consider the sparse channel with limited multipaths. When there are a large number of multipaths between the BS and the moving terminal caused by the rich reflectors, these methods cannot separate all the multipaths with a limited spatial resolution. Recently, large-scale antenna array, also known as 'massive antenna array', has gained great popularity from both academia and industry~\cite{Xie16An,Lu14An,Guo17High}. The large-scale antenna array can provide high-spatial resolution which is sufficient for dealing with multiple angle-related DFOs in high-mobility communications.
The authors in~\cite{Guo17High} have considered the high-mobility downlink transmission, where BS transmits signals to a high-speed railway (HSR). By exploiting the high-spatial resolution provided by a large-scale antenna array at HSR, the work in~\cite{Guo17High} proposed a systemic receiver design scheme with joint DFO and OFO estimation. The received signal is divided into multiple branches through high-resolution beamforming while each branch is affected by one DFO and thus the processing complexity is greatly reduced.

In this paper, we focus on the uplink transmission. We propose a Doppler pre-compensation scheme for high-mobility OFDM uplink with a large-scale uniform linear array (ULA). Considering that the multiple DFOs are related to different AoDs in the uplink, we propose to distinguish them in angle domain at the transmitter. Owing to the high-spatial resolution provided by the large-scale ULA, we generate multiple parallel branches through transmit beamforming. Since the transmitted signal in each branch is affected by one dominant DFO after passing over the channel, it can be easily compensated at the transmitter. The theoretical analysis for the Doppler spread of the equivalent uplink channel is also conducted. We also provide the simulation results to verify the proposed scheme.

%The rest of this paper is organized as follows. In Section II, the time-varying channel and signal models for high-mobility transmission are introduced, respectively. In Section III, a new transmission scheme for DFO compensation is developed. The Doppler spread for the equivalent uplink channel is analyzed in Section IV. Simulation results are provided in Section V. Section VI concludes this paper.

%Luckily, large-scale antenna array with sufficient number of receive antennas has drawn attention recently since it can provide high resolution in spatial domain~\cite{Zheng15Survey,Lu14An}.

%\emph{Notations}: $(\cdot)^\mathrm{H}$, $(\cdot)^\mathrm{T}$, and $(\cdot)^\ast$ represent the Hermitian, transposition, and conjugate operations, respectively. $\Re(\cdot)$, $\Im(\cdot)$, and $\|\cdot\|$ denote the real part, imaginary part, and Euclidean norm of a vector, respectively. %$|\cdot|$ denotes the absolute value of a scalar

\section{System Model}\label{sec:ULSysMod}
\begin{figure}[t!]
\centering
\includegraphics[scale=0.35]{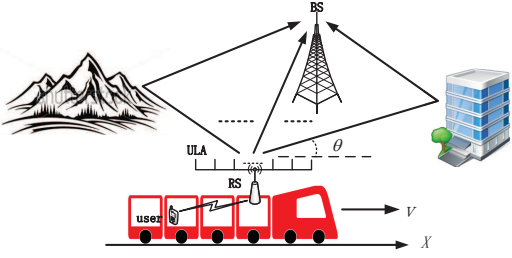}
\caption{High-mobility uplink transmission with a large-scale ULA.}
\label{fig:ULSysMod}
\vspace*{-0.15in}
\end{figure}

Consider the scenarios of high-mobility uplink communication. As in Figure~\ref{fig:ULSysMod}, a relay station (RS) with a large-scale ULA is configured on top of the train for decoding and forwarding the data between the users and the BS~\cite{Zhang11Multiple}. Since the user-to-RS link is less affected by the Doppler shifts, we only focus on the RS-to-BS link. The large-scale ULA is placed along the direction of the motion (X-axis in Figure~\ref{fig:ULSysMod}).
%In this section, we first describe the time-varying channel model, then introduce the detailed signal model.

\vspace*{-0.0in}
\subsection{Time-Varying Multipath Channel Model}\label{ssec:ULChnMod}
In the uplink transmission, the transmitted signal spreads from the RS to the BS through multipaths, which are induced by the various reflectors around the moving train, such as the buildings or mountains in Figure~\ref{fig:ULSysMod}. Assume that the large-scale ULA at the RS has $N_t$ transmit antennas and the conventional small-scale antenna array with $N_r$ receive antennas is configured at the BS. Take the channel from the $n_t$th transmit antenna to the $n_r$th receive antenna, denoted as antenna pair $\{n_r,n_t\}$, as an example. When there are rich reflectors around the moving train, the time-varying multipath channel for the antenna pair $\{n_r,n_t\}$ can be modeled as
\begin{equation}\label{eq:h_chn}
  h_{n_r,n_t}\left(n,n'\right) = \sum\limits_{l=1}^{L}{g_{n_r,n_t}(l,n-d_l)\delta\left(n'-d_l\right)},
\end{equation}
where $L$ is the total number of channel taps with different delays, $d_l$ is the relative delay of the $l$th tap, and $g_{n_r,n_t}(l,n)$ is the corresponding complex amplitude for the antenna pair $\{n_r,n_t\}$. %In a high-mobility environment, $g_{a,l}(n)$ varies with the time index $n$ due to the significant Doppler spread.
To characterize the scenarios with rich reflectors, the Jakes' channel model has been widely used in the literature~\cite{Jakes94Microwave} and an established simulator has been proposed in~\cite{Zheng03Simulation}. As in~\cite{Zheng03Simulation}, each tap is modeled as a frequency-nonselective fading channel comprised of $N_p$ propagation paths. The equivalent model for $g_{n_r,n_t}(l,n)$ is given by
\begingroup\makeatletter\def\f@size{9.5}\check@mathfonts
\def\maketag@@@#1{\hbox{\m@th\normalsize\normalfont#1}}
\begin{equation}\label{eq:h_tap}
  g_{n_r,n_t}(l,n) \!\!=\!\! \sum\limits_{q=1}^{N_p}\!\! {\alpha_{l,q} e^{j[2\pi{f_d}n{T_s}\cos{\theta_{l,q}}+\phi_{l,q}+\psi_{n_t}(\theta_{l,q})+\psi'_{n_r}(\vartheta_{l,q})]} },
\end{equation}
\endgroup
where $\alpha_{l,q}$ and $\phi_{l,q}$ are the random path gain and phase, respectively, for the $q$th path in the $l$th tap, and $T_s$ is the sampling interval. The maximum Doppler shift, $f_d$, is defined as $f_d=v/\lambda$, where $v$ is the speed of the moving terminal, and $\lambda$ is the wavelength of carrier wave. Here, $\theta_{l,q}$ and $\vartheta_{l,q}$ are the AoD and AoA, respectively, of the $q$th path in the $l$th tap relative to the direction of motion, and they are random distributed between 0 and $\pi$ in the three dimensional space. The DFO for the $q$th path in the $l$th tap is determined by the AoD and the maximum DFO, that is, $f_{l,q}={f_d}\cos{\theta_{l,q}}$.
The phase shift induced at the $n_t$th transmit antenna is denoted as $\psi_{n_t}(\theta_{l,q})$ in~\eqref{eq:h_tap}. It is determined by the antenna structure, position, and the AoD, $\theta_{l,q}$. Take the 1st transmit antenna as a reference, the phase shift, $\psi_{n_t}(\theta_{l,q})$, for the ULA can be expressed as
\begin{equation}
  \psi_{n_t}(\theta_{l,q})=2\pi(n_t-1)d_t\cos(\theta_{l,q})/\lambda,
\end{equation}
where $d_t$ is the transmit antenna element spacing. We can further determine the steering vector for the whole transmit antenna array at the AoD $\theta_{l,q}$ as
\begin{equation}
  \mathbf{a}_t(\theta_{l,q})=[e^{j\psi_1(\theta_{l,q})},\cdots,e^{j\psi_{N_t}(\theta_{l,q})}]^\mathrm{T}.
\end{equation}

Similarly, take the 1st receive antenna as a reference, the phase shift for the $n_r$th receive antenna at the AoA, $\vartheta_{l,q}$, is defined as $\psi'_{n_r}(\vartheta_{l,q})$ in~\eqref{eq:h_tap}.
%\begin{equation}
%  \psi'_{n_r}(\vartheta_{l,q})=2\pi(n_r-1)d_r\cos(\vartheta_{l,q})/\lambda,
%\end{equation}
%where $d_r$ is the receive antenna element spacing. The steering vector for the receive antenna array at the AoA, $\vartheta_{l,q}$, is written as $\mathbf{a}_r(\vartheta_{l,q})=[e^{j\psi'_1(\vartheta_{l,q})},\cdots,e^{j\psi'_{N_r}(\vartheta_{l,q})}]^\mathrm{T}$.

In a high-mobility environment, $g_{n_r,n_t}(l,n)$ varies with the time index $n$ due to the significant Doppler shifts.
Each path in~\eqref{eq:h_tap} has independent attenuation, phase, AoD, AoA and also DFO. During the transmitting period of one OFDM frame, there is little change in the position and speed of the moving terminal. Therefore, we can assume that $\alpha_{l,q}$, $\theta_{l,q}$, $\vartheta_{l,q}$, and $f_d$ are constant over the observed data frame, and only vary among different frames.
When there are rich scatters around the moving terminal, $N_p$ tends to be vary large to reflect the classical Jakes' channel model~\cite{Jakes94Microwave}, while when there are few dominant multipaths from the moving terminal to the BS,~\eqref{eq:h_chn} is simplified to the sparse channel model in~\cite{Guo13Multiple}. %Thus, the channel model discussed in this paper reflects various scenarios of the practical high-mobility transmission. %In the following, we focus on the classical Jakes' channel model.

\vspace*{-0.0in}
\subsection{Signal Model}\label{ssec:ULSigMod}
Consider the frame structure in an OFDM system, where each frame consists of $N_b$ OFDM blocks.
Denote $\mathbf{x}_m=\left[ x_{m,0},x_{m,1},\cdots ,x_{m,N_c-1} \right]^\mathrm{T}$ as the information symbols in the $m$th OFDM block, where $N_c$ is the number of subcarriers. After applying an $N_c$-point inverse discrete Fourier transform (IDFT) operator and adding the cyclic prefix (CP) of length $N_{cp}$ to each block, the resulting time-domain samples in the $m$th block can be expressed as
\begin{equation}\label{eq:s_mn}
  s_m\left(n\right) =\frac{1}{\sqrt{N_c}}\sum\limits_{k=0}^{N_c-1}{x_{m,k}e^{j\frac{2\pi kn}{N_c}}},-N_{cp}\leq n\leq N_{c}-1.
\end{equation}

%Let $N_s=N_c+N_{cp}$ be the length of an OFDM block. Then the time-domain OFDM transmitted signal can be written as
%\begin{equation}\label{eq:s_n}
%  \tilde{s}\left(n\right)=\sum\limits_{m=0}^{\infty}{s_m\left( n-m{N_s} \right)}.
%\end{equation}

Assume that the transmitted signal power is equally allocated to $N_t$ transmit antennas, that is, the signal on the $n_t$th transmit antenna is $\tilde{s}_{m,n_t}(n)=s_m(n)/\sqrt{N_t}$. We further denote $\tilde{\mathbf{s}}_{m}(n)=[\tilde{s}_{m,1}(n),\cdots,\tilde{s}_{m,N_t}(n)]^\mathrm{T}$ as the $n$th time-domain sample of the transmitted signal in the $m$th OFDM block on the whole antenna array. After passing over the above time-varying multipath channel, the received signal on the $n_r$th receive antenna is the sum of the signals transmitted from the large-scale ULA at the RS. Assume perfect time synchronization at the receiver. From~\eqref{eq:h_chn} and~\eqref{eq:s_mn}, the $n$th time-domain sample in the $m$th OFDM block at the $n_r$th receive antenna can be expressed as
%\begin{align}\label{eq:y_a}
%  {y}_{n_r}(n) = \sum\limits_{n_t=1}^{N_t}\sum\limits_{l=1}^{L} g_{n_r,n_t}(l,n)\tilde{s}(n-d_l) + {z}_{n_r}(n),
%\end{align}
\begin{align}\label{eq:y_mn}
  {y}_{m,n_r}(n) &=\!\! \sum\limits_{n_t=1}^{N_t}\sum\limits_{l=1}^{L} g_{n_r,n_t}(l,mN_s+n-d_l)\tilde{s}_{m,n_t}(n-d_l) \nonumber\\ &+ {z}_{m,n_r}(n),
\end{align}
where $N_s=N_c+N_{cp}$ is the length of an OFDM block, and $z_{m,n_r}(n)$ is the corresponding time-domain sample of the complex additive white Gaussian noise (AWGN) at the $n_r$th receive antenna. %with mean zero and variance $\sigma_n^2$. is the $n$th time-domain sample of noise in the $m$th OFDM block at the $a$th receive antenna.

%Denote $\mathbf{y}(n) = [y_{1}(n),\cdots,y_{N_r}(n)]^\mathrm{T}$ and $\mathbf{z}(n)=[z_{1}(n),\cdots,z_{N_r}(n)]^\mathrm{T}$ as the vector representation of the $n$th time-domain sample of the received signal and noise, respectively, on the whole antenna array. From~\eqref{eq:y_amn}, the received signal vector $\mathbf{y}(n)$ can be expressed as
%\begin{equation}\label{eq:y_m}
%  \mathbf{y}(n) = \sum\limits_{l=1}^{L} \sum\limits_{q=1}^{N_p} \mathbf{a}_r(\vartheta_{l,q}) \beta(\theta_{l,q},n) \mathbf{a}_t^\mathrm{T}(\theta_{l,q})\mathbf{s}(n-d_l) + \mathbf{z}(n),
%\end{equation}

Denote $\mathbf{y}_{m,n_r} = [y_{m,n_r}(0),\cdots,y_{m,n_r}(N_c-1)]$ and $\mathbf{z}_{m,n_r}=[z_{m,n_r}(0),\cdots,z_{m,n_r}(N_c-1)]$ as the vector representation of the $m$th block of the received signal and noise, respectively, on the $n_r$th receive antenna after CP removal. From~\eqref{eq:y_mn}, $\mathbf{y}_{m,n_r}$ can be expressed as
\begin{equation}\label{eq:y_m}
  \mathbf{y}_{m,n_r} \!\!=\!\! \sum\limits_{l=1}^{L} \sum\limits_{q=1}^{N_p} \rho_{l,q,n_r} \mathbf{a}_t^\mathrm{T}(\theta_{l,q}) \mathbf{S}_m(d_l) \mathbf{\Phi}_m(l,q) + \mathbf{z}_{m,n_r},
\end{equation}
where $\rho_{l,q,n_r}$ relates to the channel gain and the phase shift at the receive antenna and is written as $\rho_{l,q,n_r}=\alpha_{l,q}e^{j\phi_{l,q}}e^{j\psi'_{n_r}(\vartheta_{l,q})}$. $\mathbf{S}_{m}(d_l)$ denotes the transmitted signal matrix on the large-scale ULA after the delay of $d_l$, it can be represented as $\mathbf{S}_{m}(d_l)=[\tilde{\mathbf{s}}_{m}(0-d_l), \cdots, \tilde{\mathbf{s}}_{m}(N_c-d_l-1)] \in \mathbb{C}^{N_t\times N_c}$. $\mathbf{\Phi}_m(l,q)$ represents the following $N_c \times N_c$ diagonal phase rotation matrix introduced by the DFO $f_{l,q}$
\begin{equation}
  \mathbf{\Phi}_m(l,q) = \mathrm{diag}\{\beta_{m,0}(l,q),\cdots,\beta_{m,N_c-1}(l,q)\},
\end{equation}
where $\beta_{m,n}(l,q)=e^{j2\pi f_{l,q}(m{N_s}+n-d_l){T_s}}$.

We can see from~\eqref{eq:y_m} that the received signal is affected by multiple DFOs. When there is only one path from the RS to the BS, there exists only one DFO in the received signal, thus conventional single CFO compensation techniques can be applied. However, the multipath channel with Doppler shifts makes multiple DFOs mix at the receiver. It is hard to distinguish multiple DFOs at the receiver even with multi-antenna techniques. Note that the receiver design scheme proposed in~\cite{Guo17High} is valid in the downlink where the DFOs are related to the AoAs. However, in the uplink, the BS cannot compensate multiple DFOs with this receiver scheme since the DFOs are now associated with the AoDs. In the following, we will provide an efficient solution for DFOs pre-compensation.

\section{Transmitter Design for High-Mobility OFDM Uplink}\label{sec:TxDesign}
%In this section, we propose a new transmitter design scheme for high-mobility OFDM uplink transmission. We first develop the main design procedures, then the beamforming network and DFO compensation will be introduced in detail.

\vspace*{-0.0in}
\subsection{Motivation and Transmitter Design}\label{ssec:chn_est}
Before introducing the proposed scheme, we first illustrate why it is beneficial to exploit a massive ULA at the transmitter in the uplink. The reason lies in the following three aspects:

First, multi-antenna techniques can separate multiple DFOs. As discussed previously, the received signal on a single antenna consists of multiple DFOs. Since the DFOs are associated with the AoDs, it is hard to separate them with one single antenna in either time or frequency domains. The multi-antenna techniques can provide spatial resolution thus have the potential to separate multiple DFOs in angle domain. %exploit the spatial characteristic of the channel

Second, only the transmitter can distinguish multiple DFOs in the uplink transmission. As introduced in \cite{Guo13Multiple}, in the uplink transmission, one AoA may be associated with multiple DFOs while the AoD has a one-to-one relationship with the DFO. Therefore, the receiver cannot distinguish multiple DFOs according to the AoAs while only the transmitter can exploit the AoDs to distinguish multiple DFOs.

Third, conventional small-scale antenna arrays cannot provide enough spatial resolution. Some previous work has adopted small-scale antenna arrays in high-mobility systems. They are efficient for the sparse channels~\cite{Zhang11Multiple,Guo13Multiple,Yang13Beamforming} when there are very limited multipaths between transceivers. However, when there are a large number of multipaths with different DFOs, conventional small-scale antenna arrays become powerless.
Here we should note that, large-scale antenna systems have gained much interest~\cite{Xie16An,Lu14An}, where the transmitter or the receiver are equipped with large number of antennas. %The large-scale antenna array can provide high spatial resolution to separate the large number of multipaths.
These systems can provide high-spatial resolution which is sufficient for dealing with large number of DFOs with different AoDs. %In the proposed scheme, we configure a large-scale ULA at the transmitter to separate multiple DFOs via transmit beamforming.
%as introduced in Section~\ref{ssec:ULChnMod}, the multiple DFOs are associated with the AoDs of the multipaths, thus they can only be distinguished in the spatial domain.

\begin{figure}[t!]
\centering
\includegraphics[scale=0.53]{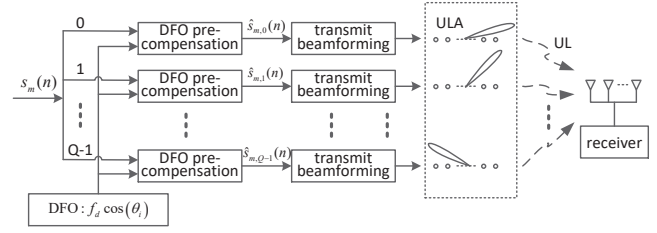}
\caption{Diagram of the transmitter design.}
\label{fig:ULTxDesign}
\vspace*{-0.15in}
\end{figure}

Based on the above observations, we propose to utilize a large-scale ULA at the transmitter to cope with the significant Doppler spread in high-speed transmission.
Figure~\ref{fig:ULTxDesign} shows the diagram of the transmitter design. The transmitted signal passes through $Q$ parallel branches simultaneously. In each branch, we first perform DFO pre-compensation to the signal, then we design a high-resolution beamforming network and transmit the signal towards the pre-selected direction through transmit beamforming. Since the transmitted signal in each branch is limited to a narrow beam, it is mainly affected by one DFO when passing over the channel. Thus it is easy to perform DFO pre-compensation at the transmitter, and it also simplifies the channel estimation and equalization at the receiver. %In the following, we introduce the details for these procedures.

\vspace*{-0.0in}
\subsection{Angle-Domain DFO Pre-Compensation}\label{ssec:bf}
%As discussed previously, the received signal on a single antenna consists of multiple DFOs. It is hard to separate multiple DFOs with one single antenna. Since the DFOs are associated with the AoDs, we will solve this problem in spatial domain. Recently, large-scale antenna systems have gained much interest~\cite{Zheng15Survey,Lu14An}, where the transmitter or the receiver are equipped with large number of antennas.
%Recently, large-scale antenna systems have gained much interest~\cite{Zheng15Survey,Lu14An}, where the transmitter or the receiver are equipped with sufficient number of antennas, e.g., 100 or more.
%These systems can provide high spatial resolution which is sufficient for dealing with large number of DFOs with different AoDs. In the proposed scheme, we configure a large-scale ULA at the transmitter to separate multiple DFOs via transmit beamforming.

As introduced in~\cite{Guo17High}, we place the large-scale ULA in the direction of the motion to make the distribution of the beam pattern in accordance with that of the DFOs. The antenna element spacing is selected as $d_t<\lambda/2$ to produce only one main beam in each transmit beamforming. We further assume that $Q$ different values of $\theta$ are evenly selected between $0^\circ$ and $180^\circ$ to design the beamforming network, which are denoted by $\theta_i, i=0,1,\cdots,Q-1$. Note that the beamforming is performed in the whole range without estimating the AoDs of the multipaths, and the beamforming network is designed previously which reduces the complexity. This is quite different from the existing competitors~\cite{Zhang11Multiple,Yang13Beamforming,Guo13Multiple}.

To transmit signal towards only one direction, the goal of beamforming is to maintain the signal transmitted to the desired direction while suppressing the signals leaked to other directions. It can be easily implemented through the match filter beamformer. The beamforming weight vector for direction $\theta_i$ is determined by the steering vector, that is,
\begin{equation}\label{eq:w}
  \mathbf{w}(\theta_i) = \frac{\mathbf{a}_t(\theta_i)e^{j\phi'(\theta_i)}}{\left\|\sum_{i=0}^{Q-1}\mathbf{a}_t(\theta_i)e^{j\phi'(\theta_i)}\right\|} = \eta\mathbf{a}_t(\theta_i)e^{j\phi'(\theta_i)},
\end{equation}
where $\eta$ is the normalized parameter that used to restrict the transmitted signal power on multiple antennas and $\phi'(\theta_i)$ is a random phase introduced at the direction $\theta_i$. It is used to make the channel auto-correlation function independent of time and will be discussed in next section. Since the transmitted signal in each branch is affected by one dominant DFO after passing over the channel, we can easily compensate the DFO at the transmitter previously. Assume that the maximum DFO is known at the transmitter. For the $i$th branch, the DFO induced by the channel is $f_d\theta_i$ after transmit beamforming towards $\theta_i$. Thus, the signal after DFO pre-compensation is determined as
\begin{equation}
  \hat{s}_{m,i}(n) = {s}_{m}(n) e^{-j2\pi f_{d}\cos\theta_i(m{N_s}+n){T_s}}.
\end{equation}

After transmit beamforming towards $\theta_i$, the transmitted signal vector on the whole antenna array is given by
\begin{equation}
  \hat{\mathbf{s}}_{m,i}(n) = \mathbf{w}^*(\theta_i)\hat{s}_{m,i}(n).
\end{equation}

Similar to $\mathbf{S}_m(d_l)$ in~\eqref{eq:y_m}, define $\hat{\mathbf{S}}_{m,i}(d_l)$ as the transmitted signal matrix in the $m$th OFDM block of the $i$th branch after the delay of $d_l$, and it can be expressed as
\begin{equation}\label{eq:S_mi_1}
  \hat{\mathbf{S}}_{m,i}(d_l) = [\hat{\mathbf{s}}_{m,i}(0-d_l), \cdots, \hat{\mathbf{s}}_{m,i}(N_c-d_l-1)].
\end{equation}

We further rewrite~\eqref{eq:S_mi_1} in the following matrix form
\begin{equation}\label{eq:S_mi_2}
  \hat{\mathbf{S}}_{m,i}(d_l) = \mathbf{w}^*(\theta_i)\mathbf{s}_m(d_l)\mathbf{\Psi}_{m,i}(d_l),
\end{equation}
where $\mathbf{s}_{m}(d_l)=[s_{m}(0-d_l), \cdots, s_{m}(N_c-d_l-1)]$ is the initial transmitted signal vector after the delay of $d_l$, and $\mathbf{\Psi}_{m,i}(d_l)$ is the DFO pre-compensation matrix in the $i$th branch and can be expressed as
\begin{equation}
  \mathbf{\Psi}_{m,i}(d_l) = \mathrm{diag}\{\hat{\beta}_{m,0,i}(d_l),\cdots,\hat{\beta}_{m,N_c-1,i}(d_l)\},
\end{equation}
where $\hat{\beta}_{m,n,i}(d_l)=e^{-j2\pi f_d\cos\theta_i(m{N_s}+n-d_l){T_s}}$.

After replacing $\mathbf{S}_{m}(d_l)$ in~\eqref{eq:y_m} by $\hat{\mathbf{S}}_{m,i}(d_l)$, the received signal vector from the $i$th branch can be written as
\begingroup\makeatletter\def\f@size{9.5}\check@mathfonts
\def\maketag@@@#1{\hbox{\m@th\normalsize\normalfont#1}}%
\begin{equation}\label{eq:r_mi}
  \mathbf{r}_{m,n_r,i} \!\!=\!\! \sum\limits_{l=1}^{L} \sum\limits_{q=1}^{N_p} \rho_{l,q,n_r} \mathbf{a}_t^\mathrm{T}(\theta_{l,q}) \hat{\mathbf{S}}_{m,i}(d_l) \mathbf{\Phi}_m(l,q) + \mathbf{z}_{m,n_r}.
\end{equation}
\endgroup

From~\eqref{eq:S_mi_2}, we can further divide~\eqref{eq:r_mi} into~\eqref{eq:r_mi_1}, located at the top of the following page.
\begin{figure*}[t]
\begin{align*}\label{eq:r_mi_1}
  \mathbf{r}_{m,n_r,i}= \underbrace{\!\!\!\sum\limits_{l,q,\theta_{l,q}=\theta_i}\!\!\! \rho_{l,q,n_r}\eta{N_t}e^{-j\phi'(\theta_i)}\mathbf{s}_m(d_l)}_{desired\ signal} + \underbrace{\sum\limits_{l',q',\theta_{l',q'}\neq\theta_i}\!\!\! \rho_{l',q',n_r}\mathbf{w}^\mathrm{H}(\theta_i) \mathbf{a}_t(\theta_{l',q'}) \mathbf{s}_m(d_{l'})\mathbf{\Psi}_{m,i}(d_{l'})\mathbf{\Phi}_m(l',q')}_{interference} + \underbrace{\mathbf{z}_{m,n_r}}_{noise}
\tag{16} % 设置公式编号为14
\end{align*}
\hrulefill % 在公式之后加上一条分割线，长度为页面宽度
\vspace*{-0.1in}
\end{figure*}
%\begingroup\makeatletter\def\f@size{9}\check@mathfonts
%\def\maketag@@@#1{\hbox{\m@th\normalsize\normalfont#1}}%
%\begin{align}\label{eq:r_m_1}
%  &\mathbf{r}_{m,n_r,i}= \underbrace{\sum\limits_{l,q,\theta_{l,q}=\theta_i} \rho_{l,q,n_r}\eta{N_t}\mathbf{s}_m(d_l)}_{desired\ signal} \nonumber\\
%  &+\!\!\!\! \underbrace{\sum\limits_{l',q',\theta_{l',q'}\neq\theta_i}\!\!\!\! \rho_{l',q',n_r}\mathbf{w}^\mathrm{H}(\theta_i) \mathbf{a}(\theta_{l',q'}) \mathbf{s}_m(d_{l'})\mathbf{\Psi}_{m,i}(d_{l'})\mathbf{\Phi}_m(l',q')}_{interference} \nonumber\\
%  &+ \underbrace{\mathbf{z}_{m,n_r}}_{noise},
%\end{align}\endgroup

In~\eqref{eq:r_mi_1}, the first term represents the desired signal from the AoD $\theta_i$ which is the same as the standard time-invariant received signal model under frequency-selective channel. The second term is the interference transmitted from other directions and is affected by the residual DFOs. The third term is the noise at the $n_r$th receive antenna. When the number of transmit antennas $N_t$ is sufficiently large, $\mathbf{w}^\mathrm{H}(\theta_i)\mathbf{a}_t(\theta_{l',q'})\simeq 0$ holds for $\theta_i\neq\theta_{l',q'}$. Therefore, the second term is greatly suppressed through the high-resolution transmit beamforming.% techniques. %When there is no path with the AoD $\theta_i$, the resultant signal in (16) after transmit beamforming is comprised of only weak interference and noise.

Finally, the received signal at the $n_r$th receive antenna is the sum of signals from $Q$ parallel branches, that is%transmitted
\setcounter{equation}{16}
\begin{equation}\label{eq:r_m_ant}
  \tilde{\mathbf{r}}_{m,n_r} = \sum\limits_{i=0}^{Q-1}\mathbf{r}_{m,n_r,i}.
\end{equation}

Through the above analysis, in each beamforming branch, the dominant DFO caused by the high-mobility of the train has been compensated previously. Now the uplink channel can be considered as time-invariant approximately when the interference can be mitigated to a tolerable magnitude. Especially, the interference turns to zero when $N_t$ approaches to infinite, then the channel becomes exactly time-invariant. The conventional channel estimation methods can be carried out to estimate the channel response in each receive antenna~\cite{Zhang11Multiple,Yang13Beamforming}. Then, the maximum-ratio-combining (MRC) is utilized to detect the transmitted data through multiple receive antennas.

\section{Doppler Spread Analysis}\label{sec:DS}
%In~\cite{Guo17High}, the downlink transmission with large-scale antenna array has been investigated, where each beamforming branch is considered as independent at the receiver through receive beamforming. However, in the proposed scheme,
As introduced above, multiple transmit beamforming branches are combined together at the receiver. Through DFO pre-compensation and transmit beamforming, the dominant DFO in each branch is eliminated. When the number of transmit antennas is sufficiently large, the time-variation of channel can be neglected. However, the number of transmit antennas is limited in practice which means that the residual DFOs for all the branches cannot be ignored.
In this section, we theoretically analyze the effect of the residual Doppler shifts on the proposed scheme. As in~\cite{Souden09Robust}, the Doppler spread is used to evaluate the channel variation.

Since the channel taps are independent and have the same statistical property, we analyze the Doppler spread for only one tap for simplicity, that is, the Jakes' channel is comprised of single tap with delay $d_1=0$. Considering the reference antenna at the receiver, after omitting the receive antenna index $n_r$ and the noise term, the received signal in~\eqref{eq:r_m_ant} turns to
\begingroup\makeatletter\def\f@size{9.2}\check@mathfonts
\def\maketag@@@#1{\hbox{\m@th\normalsize\normalfont#1}}%
\begin{equation}\label{eq:r_m}
  \tilde{\mathbf{r}}_{m} \!\!=\!\! \sum\limits_{i=0}^{Q-1}\sum\limits_{q=1}^{N_p} \rho_{1,q} \mathbf{w}^\mathrm{H}(\theta_i) \mathbf{a}_t(\theta_{1,q}) \mathbf{s}_m(d_1)\mathbf{\Psi}_{m,i}(d_1) \mathbf{\Phi}_m(1,q).
\end{equation}
\endgroup

Assume that each tap has a large number of multipaths with the AoDs between $0$ and $\pi$ and transmit beamforming is performed continuously between $0$ and $\pi$. Then the summations in~\eqref{eq:r_m} can be replaced by the integrations. After replacing $\theta_{1,q}$ and $\theta_{i}$ by $\tilde{\theta}$ and $\theta$, respectively, the equivalent uplink channel can be expressed in the continuous-time form as
\begin{equation}\label{eq:g_chn_bf}
\tilde{g}(t) \!=\! E_0 \int_0^{\pi} \int_0^{\pi}\!\! \alpha(\tilde{\theta})G(\theta,\tilde{\theta}) e^{j(\omega_d t y(\theta,\tilde{\theta})+\varphi(\theta,\tilde{\theta}) )} d\tilde{\theta} d\theta,
\end{equation}
where $t$ is the continuous-time index, $E_0$ is a scaling constant, $\alpha(\tilde{\theta})$ is the random gain for the path with AoD $\tilde{\theta}$, $\omega_d=2\pi f_d$, and $y(\theta,\tilde{\theta})=\cos{\tilde{\theta}}-\cos{\theta}$. 
Here, $G(\theta,\tilde{\theta})$ is the antenna gain at the direction $\tilde{\theta}$ when applying the match filter beamformer towards the direction $\theta$, and is determined as
\begin{equation}\label{eq:Gain}
  G(\theta,\tilde{\theta}) = \frac{1}{M}\mathbf{a}_t^\mathrm{H}(\theta)\mathbf{a}_t(\tilde{\theta})=\frac{\sin[\frac{\pi Md}{\lambda}y(\theta,\tilde{\theta})]} {M\sin[\frac{\pi d}{\lambda}y(\theta,\tilde{\theta})]}.
\end{equation}
%Note that the difference between the weight vectors used in~\eqref{eq:w} and~\eqref{eq:Gain} is a constant coefficient which will not affect the following analysis.

In~\eqref{eq:g_chn_bf}, the random phase $\varphi(\theta,\tilde{\theta})$ is expressed as $\varphi(\theta,\tilde{\theta})=\phi(\tilde{\theta})+\phi'(\theta)$, where $\phi'(\theta)$ is the random phase introduced at the beamforming direction $\theta$ at the transmitter and $\phi(\tilde{\theta})$ is the random phase for the path with AoD $\tilde{\theta}$. 
We introduce the random phase $\phi(\tilde{\theta})$ in the beamforming network such that the equivalent uplink channel satisfies stationary distributed. Note that almost all the existing channel estimation methods are based on the assumption of stationary channel.
Since $\phi(\tilde{\theta})$ and $\phi'(\theta)$ are statistically independent and are randomly selected between $0$ and $2\pi$, we can easily prove that the autocorrelation for the equivalent fading channel in~\eqref{eq:g_chn_bf} is independent of the time index $t$ and can be expressed as
\begingroup\makeatletter\def\f@size{9.2}\check@mathfonts
\def\maketag@@@#1{\hbox{\m@th\normalsize\normalfont#1}}%
\begin{equation}\label{eq:R}
R_{\tilde{g}\tilde{g}}(\tau) \!=\! E_0^2 \int_0^{\pi} \int_0^{\pi}\!\! \mathrm{E}\{\alpha^2(\tilde{\theta})\} |G(\theta,\tilde{\theta})|^2 e^{-j\omega_d \tau y(\theta,\tilde{\theta})} d\tilde{\theta} d\theta.
\end{equation}
\endgroup

As in~\cite{Zheng03Simulation}, we assume $\int_0^{\pi}\mathrm{E}\{C^2(\tilde{\theta})\}d\tilde{\theta}=1$ and $E_0=\sqrt{2}$. Now there exists a constant coefficient between~\eqref{eq:R} and the accurate channel autocorrelation which will not affect the following analysis.
The power spectrum density (PSD) of the channel is defined as
\begin{equation}\label{eq:PSD_def}
P(\omega) = \int_{-\infty}^{+\infty} R_{\tilde{g}\tilde{g}}(\tau)e^{-j\omega\tau} d\tau.
\end{equation}

\begin{lemma} \label{Lemma:PSD}
The PSD of the equivalent uplink channel can be approximated as
\begingroup\makeatletter\def\f@size{9.7}\check@mathfonts
\def\maketag@@@#1{\hbox{\m@th\normalsize\normalfont#1}}%
\begin{align}\label{eq:PSD}
P(\omega) &\simeq \!\!\!\!\sum_{\substack{I_{min}\leq i\leq I_{max} \\ i\neq\{-1,0\}}}\!\!\!\!\!\!\! 2{D_i}\bar{C}_i(1+\cos{2(\omega+W_i)t_0})\bar{X}(j(\omega+W_i)) \nonumber\\
  & + C_0\delta(\omega)+2C_1(1+\cos\omega t_0)X(j\omega),
\end{align}
\endgroup
where $C_0=\frac{8{\Delta_m\theta_0}}{\pi}$ with $\Delta_m=\arccos(1-\frac{\lambda}{Md})$ and $\theta_0=\arcsin(\frac{\lambda}{Md{\Delta_m}})$, $t_0=\frac{\pi Md}{\omega_d\lambda}$, $C_1 = -\frac{2}{\omega_d}\ln\tan\frac{\theta_0}{2}$, $D_i=\left|\frac{1}{M\sin[\frac{(2i+1)\pi}{2M}]}\right|^2$, and $W_i=\frac{(2i+1)\lambda}{2Md}\omega_d$. Here, $\bar{C}_i$ is determined by the following function
\begin{equation*}
\bar{C}_i = \left\{\! \begin{array}{ll} \frac{1}{\omega_d} \int_{\bar{\theta}_0}^{b_{i}}  \frac{1}{\sqrt{1-(\cos\theta-\frac{(2i+1)\lambda}{2Md})^2}}d\theta, & \mathrm{if}\ i>0 \\ \frac{1}{\omega_d} \int_{a_i}^{\pi-\bar{\theta}_0} \frac{1}{\sqrt{1-(\cos\theta-\frac{(2i+1)\lambda}{2Md})^2}}d\theta, & \mathrm{if}\ i<-1 \end{array} \right.
\end{equation*}
where $\bar{\theta}_0=\arcsin( \frac{\lambda}{2Md\arccos(1-\frac{\lambda}{2Md})} )$, $a_i=\arccos(1+\frac{(2i+1)\lambda}{2Md})$, and $b_i=\arccos(-1+\frac{(2i+1)\lambda}{2Md})$. In~\eqref{eq:PSD}, $\delta(\omega)$ is the impulse-response function, and $X(j\omega)$ and $\bar{X}(j\omega)$ are the following rectangular window functions %with $W_0=\frac{\omega_d\lambda}{Md}$ and $\bar{W}_0=W_0/2$, respectively
\begin{equation*}
X(j\omega) = \left\{\! \begin{array}{ll} 1, & |\omega|<W_0 \\ 0, & |\omega|>W_0 \end{array} \right.
\end{equation*}
\begin{equation*}
\bar{X}(j\omega) = \left\{\! \begin{array}{ll} 1, & |\omega|<\bar{W}_0 \\ 0, & |\omega|>\bar{W}_0 \end{array} \right.
\end{equation*}
where $W_0=\frac{\omega_d\lambda}{Md}$ and $\bar{W}_0=W_0/2$. The range for the integer $i$ is determined by the range of the antenna sidelobes and is given by $\{i|i\in[I_{min},I_{max}]\ \mathrm{and}\ i\neq0,-1\}$, where $I_{max}=\lfloor\frac{2Md}{\lambda}-\frac{1}{2}\rfloor$ and $I_{min}=\lceil-\frac{2Md}{\lambda}-\frac{1}{2}\rceil$.
\end{lemma}

The detailed derivation is omitted here due to space limitation. The Doppler spread is defined as~\cite{Souden09Robust}
\begin{equation}\label{eq:DS_def}
  \sigma_{DS} = \left( \frac{\int_{-2\omega_d}^{2\omega_d}\omega^2P(\omega)d\omega}{\int_{-2\omega_d}^{2\omega_d}P(\omega)d\omega} \right)^{\frac{1}{2}}.
\end{equation}

After substituting~\eqref{eq:PSD} into~\eqref{eq:DS_def}, we have
\begin{lemma} \label{Lemma:PSD}
The Doppler spread of the equivalent uplink channel is expressed as
\begin{equation}\label{eq:DS}
  \sigma_{DS} = \left( \Gamma/\Lambda \right)^{\frac{1}{2}},
\end{equation}
where
\begin{equation*}
  \Lambda = C_0 + 4{C_1}{W_0} + \!\!\sum_{\substack{I_{min}\leq i\leq I_{max} \\ i\neq\{-1,0\}}}\!\!\! 4{D_i}{\bar{C}_i}{\bar{W}_0},
\end{equation*}
and
\begingroup\makeatletter\def\f@size{9.7}\check@mathfonts
\def\maketag@@@#1{\hbox{\m@th\normalsize\normalfont#1}}%
\begin{align*}
\Gamma &=\frac{4{C_1}{W_0^3}}{3}-\frac{8{C_1}{W_0}}{t_0^2} \nonumber\\
&+ \!\!\sum_{\substack{I_{min}\leq i\leq I_{max} \\ i\neq\{-1,0\}}}\!\!\! \left[ \frac{4{D_i}{\bar{C}_i}{\bar{W}_0^3}}{3}-\frac{2{D_i}{\bar{C}_i}{\bar{W}_0}}{{t}_0^2}+4{D_i}{\bar{C}_i}{\bar{W}_0}{W_i^2}\right].
\end{align*}
\endgroup
\end{lemma}

The detailed derivation is omitted here due to space limitation. We can further rewrite the Doppler spread in~\eqref{eq:DS} as
\begin{equation}
\sigma_{DS} = \kappa\omega_d,
\end{equation}
where $\kappa$ is a coefficient irrelevant to the maximum DFO $\omega_d$. We can see that the Doppler spread of the equivalent channel is the linear function of $\omega_d$.
%The detailed derivation is omitted here due to the space limitation.
Note that the Doppler spread for the Jakes' channel model is also a linear function of $\omega_d$~\cite{Souden09Robust} and is expressed as $\sigma_{J} = \omega_d/\sqrt{2}$.
%\begin{equation}
%\sigma_{J} = \omega_d/\sqrt{2}.
%\end{equation}

\section{Simulation Results}
In this section, we evaluate the performance of the proposed transmission scheme through numerical simulations. We consider the frame structure in an OFDM system, where the first block in a frame is the training block and the remaining blocks are used for transmitting data symbols to the receiver.
We assume the Jakes' channel model between the BS and the moving terminal as in~\cite{Zheng03Simulation}.
The transmit antenna element spacing is $d_t=0.45\lambda$ and the beamforming network is designed with the interval of $2^\circ$. The other simulation parameters are shown in Table~\ref{table:SimuPara-UL}. % for angles between $0^\circ$ and $180^\circ$ with the interval of $1^\circ$.

\begin{table}[t!]
	\renewcommand{\arraystretch}{1.2}
	\footnotesize
	\caption{Simulation Parameters\vspace*{-0.0in}}%
	\centering % centering table
	\label{table:SimuPara-UL}
	\begin{tabular}{|l|l|}
		\hline number of subcarriers & 128\\
		\hline carrier frequency & 3GHz\\
		\hline wavelength of carrier wave & $\lambda=0.1\mathrm{m}$ \\
		\hline number of blocks in each frame & 5\\
		\hline duration of each block & $T_b=0.1\mathrm{ms}$\\
		\hline maximum DFO & $f_d=1\mathrm{KHz}$ for 360km/h\\
		\hline normalized maximum DFO & $f_d T_b=0.1$\\
		\hline modulation type & 16QAM\\
		\hline \multirow{2}{*}{antenna configuration} & ULA at both Tx and Rx \\
		\cline{2-2}
		&Tx: 128/256/512/1024, Rx: 4\\
		\hline  \multirow{3}{*} {channel parameter}
		& tap number: 6\\
		\cline{2-2}
		& path in each tap: 64\\
		\cline{2-2}
		& maximum channel delay: 16 \\
%        \cline{2-2}
%        & AoD: $0^\circ\sim360^\circ$ AoA: $60^\circ\pm30^\circ$ \\
		\hline uplink receiver type & MRC-LS \\
		%\hline UE power & 23dBm maximum transmit power \\
		%\hline Min. distance from UE to BS	& $\geq$ 35m\\
		%\hline Thermal noise spectral density & -174dBm$/$Hz \\
		\hline
	\end{tabular}
	\vspace*{-0.1in}
\end{table}

\begin{figure}[t]
\centering
\includegraphics[scale=0.45]{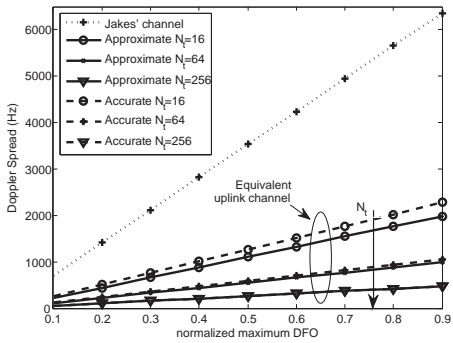}
\caption{The Doppler spread of the equivalent uplink channel.}
\label{fig:DS_fd}
\vspace*{-0.15in}
\end{figure}

First, we verify the Doppler spread analysis for the equivalent uplink channel in Section~\ref{sec:DS}. %In Figure~\ref{fig:PSD}, we compare the PSD of the equivalent channel expressed in~\eqref{eq:PSD} with the Jakes' channel when $f_d=1\mathrm{KHz}$. The results demonstrate that the dominant PSD concentrates around the frequency of zero after transmit processing while the PSD is around the maximum DFO in the Jakes' channel model. We can also see that the range for PSD is doubled in our proposed scheme which is caused by the DFO pre-compensation.
Figure~\ref{fig:DS_fd} shows the Doppler spread of the equivalent channel under different values of maximum DFO $f_d$ and transmit antenna number $N_t$. The accurate Doppler spread in dashed curves is calculated by using the definitions in~\eqref{eq:PSD_def} and~\eqref{eq:DS_def} while the analytical approximation in solid curves is obtained by using~\eqref{eq:PSD} and~\eqref{eq:DS}. For comparison, we also include the Doppler spread for the Jakes' channel. It is clear that the Doppler spread is significantly reduced after the proposed transmit processing. We also make the following observations: The approximation of Doppler spread gets closer to the accurate ones when $N_t$ gets larger. The Doppler spread of the equivalent channel is the linear function of $\omega_d$ as discussed in Section~\ref{sec:DS} and the slope is determined by $N_t$. With the increase of $N_t$, the slope of Doppler spread reduces which means that the time-variation of channel is mitigated more obviously.
%Figure () further shows the Doppler spread of the equivalent channel under different numbers of transmit antennas $N_t$. We can observe that the Doppler spread reduces with the increase of $N_t$ and changes slowly when a large number of antennas are configured at the transmitter.

%\begin{figure}[t]
%\centering
%\includegraphics[scale=0.45]{figures/PSD_v1}
%\caption{The PSD of the equivalent uplink channel.}
%\label{fig:PSD}
%\vspace*{-0.1in}
%\end{figure}

\begin{figure}[t]
\centering
\includegraphics[scale=0.45]{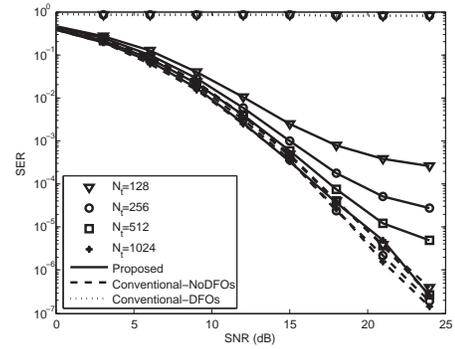}
\caption{The SER performance of the proposed scheme.}
\label{fig:SER_SNR}
\vspace*{-0.15in}
\end{figure}

Next, we evaluate the symbol error rate (SER) performance of our proposed scheme under different values of $N_t$ in Figure~\ref{fig:SER_SNR}. For comparison, the conventional transmission schemes with and without DFOs are also included as the benchmark, called Conventional-DFOs and Conventional-NoDFOs, respectively. In conventional transmission, there is no Doppler compensation at both transmitter and receiver and the signal is transmitted through multiple beamforming as in our proposed scheme. Moreover, the conventional time-invariant channel estimation based on least square (LS) is performed at the receiver in all schemes.
The results demonstrate the effectiveness of the proposed scheme. Especially, our scheme outperforms Conventional-DFOs dramatically since Conventional-DFOs suffers from high Doppler spread. From this figure, the SER performance of the proposed scheme gets closer to Conventional-NoDFOs when more antennas are configured on the RS. We can conclude that when the number of transmit antennas is sufficiently large, the BS can definitely neglect the time-variation of channel and thus exploit the conventional channel estimation and equalization methods.
%Therefore, the simulation results indicate that our proposed scheme can significantly overcome the SER degradation caused by the time-varying multipath channel with multiple CFOs.

\section{Conclusion}
In this paper, we investigate the high-mobility OFDM uplink transmission when there exists multiple DFOs. We configure a massive ULA at the transmitter and adopt the DFO pre-compensation and transmit beamforming. Then the equivalent uplink channel can be considered as time-invariant and the conventional channel estimation methods are directly used to recover the transmitted data. Both the theoretical and simulation results are provided to validate the proposed scheme. % at the receiver

\bibliographystyle{IEEEtran}
\bibliography{References}

% that's all folks
\end{document}